 \documentclass[aps,prl,superscriptaddress,showpacs,floatfix]{revtex4}

\usepackage{graphicx}
\usepackage{dcolumn}
\usepackage{bm}
\usepackage[mathlines]{lineno}

\usepackage{color}%

\begin{document}

\title{\hfill {\small Phys. Rev. Lett. {\bf 108} (2012)}\\
       Designing electrical contacts to MoS$_2$ monolayers:
       A computational study}

\author{Igor Popov}
\affiliation{School of Physics and CRANN, Trinity College
             Dublin 2, Ireland}

\author{Gotthard Seifert}
\affiliation{Physikalische Chemie, Technische Universit\"{a}t Dresden,
             D-01062 Dresden, Germany}

\author{David Tom\'{a}nek}
\email[E-mail: ]{tomanek@pa.msu.edu}%
\affiliation{Physics and Astronomy Department,
             Michigan State University,
             East Lansing, Michigan 48824-2320, USA}

\date{\today}


\begin{abstract}
Studying the reason, why single-layer molybdenum disulfide
(MoS$_2$)
appears to fall
short of its promising potential in flexible nanoelectronics, we
found that the nature of contacts plays a more important role than
the semiconductor itself. In order to understand the nature of
MoS$_2$/metal contacts, we performed \textit{ab initio} density
functional theory calculations for the geometry, bonding and
electronic structure of the contact region. We found that the most
common contact metal (Au) is rather inefficient for electron
injection into single-layer MoS$_2$ and propose Ti as a
representative example of suitable alternative electrode
materials.
\end{abstract}



\maketitle

Contrary to popular perception, contacts often play a more crucial
role in nanoscale electronics than the semiconducting material
itself\cite{Leonard2006,Nanocontact2008}. Whereas contacts in
Si-based devices are no longer considered a problem after many
decades of optimization,
engineering optimum contacts to electronic nano-devices consisting
of silicon\cite{LandmanPRL2000} or carbon (e.g. nanotubes or
graphene)\cite{Leonard2006,DT178} has become a major challenge for
the field.
More recently, the layered molybdenum disulphide (MoS$_2$)
compound, which is structurally very flexible, has emerged as a
promising alternative to silicon-based, carbon-based, and
molecular electronics\cite{Kis,Yoon-MoS2-2011}. Bulk MoS$_2$, a
well-established low-cost lubricant, has an indirect band gap of
1.2~eV\cite{MoS2-bandgap} and a rather high carrier
mobility\cite{Mooser1967,Podzorov}. In contrast to the bulk
material, the observed electron mobility in single-layer MoS$_2$
is unexpectedly low\cite{Kis,Novoselov2005}.

Here we
propose that the observed low electron mobility in MoS$_2$ may not
represent an intrinsic property of the semiconducting single
layer, but was possibly biased by unfavorable contacts, which can
dominate the electronic characteristics of MoS$_2$-based
nano-electronic devices.
Our \textit{ab initio} density functional theory (DFT)
calculations for the electronic structure of MoS$_2$/metal
contacts indicate that Au, the most common contact metal in this
system\cite{Kis}, forms a tunnel barrier at the interface, which
suppresses electron injection into MoS$_2$.
This is possibly the true reason, why the observed carrier
mobility in single-layer MoS$_2$ is lower than
expected\cite{Kis,Novoselov2005}.

Searching for better contacts than provided by Au, we focussed on
metals with a low work function that would efficiently inject
electrons into the conduction band of MoS$_2$. Among transition
metals with $d$ orbitals that may favorably mix with the Mo$4d$
states, we identified Sc, Ti and Zr as suitable candidates. Among
these, Sc and Zr are less suitable due to a large lattice
mismatch, and Ti emerges as an ideal candidate with only 1\%
mismatch to MoS$_2$. As we will show, the MoS$_2$/Ti interface
displays a much higher density of delocalized states at $E_F$ than
the MoS$_2$/Au contact. Similar to Au, Ti fulfills also other
criteria required of a good contact material in electronics, such
as high conductivity and chemical, thermal and electrical
stability. Therefore, Ti is being used widely as a contact metal
in modern microelectronics.

Our DFT calculations use the Perdew-Burke-Ernzerhof form of the
exchange-correlation functional\cite{PBE}, as implemented in the
SIESTA code\cite{Soler}. A similar approach had been successfully
used to characterize transition metal chalcogenide
nanowires\cite{Popov_PRL,Yang_PRL} and their contacts to metal
electrodes\cite{Popov_APL}.
The behavior of valence electrons was described by norm-conserving
Troullier-Martins pseudopotentials \cite{Troullier91} with partial
core corrections.
We used a double-zeta basis, including initially unoccupied Mo$5p$
orbitals. The Brillouin zone of the periodic array of
MoS$_2$/metal slabs, separated by a 33~{\AA} vacuum region, was
sampled by a $8{\times}16{\times}1$ $k-$point grid. We limited the
range of the localized orbitals in such a way that the energy
shift caused by their spatial confinement was no more than
140~meV\cite{SIESTA_PAO}. The charge density and potentials were
determined on a real-space grid with a mesh cutoff energy of
$200$~Ry, which was sufficient to achieve a total energy
convergence of better than $0.1$~meV/unit cell during the
self-consistency iterations.

\begin{figure}[tbp!]
\includegraphics[width=0.9\columnwidth]{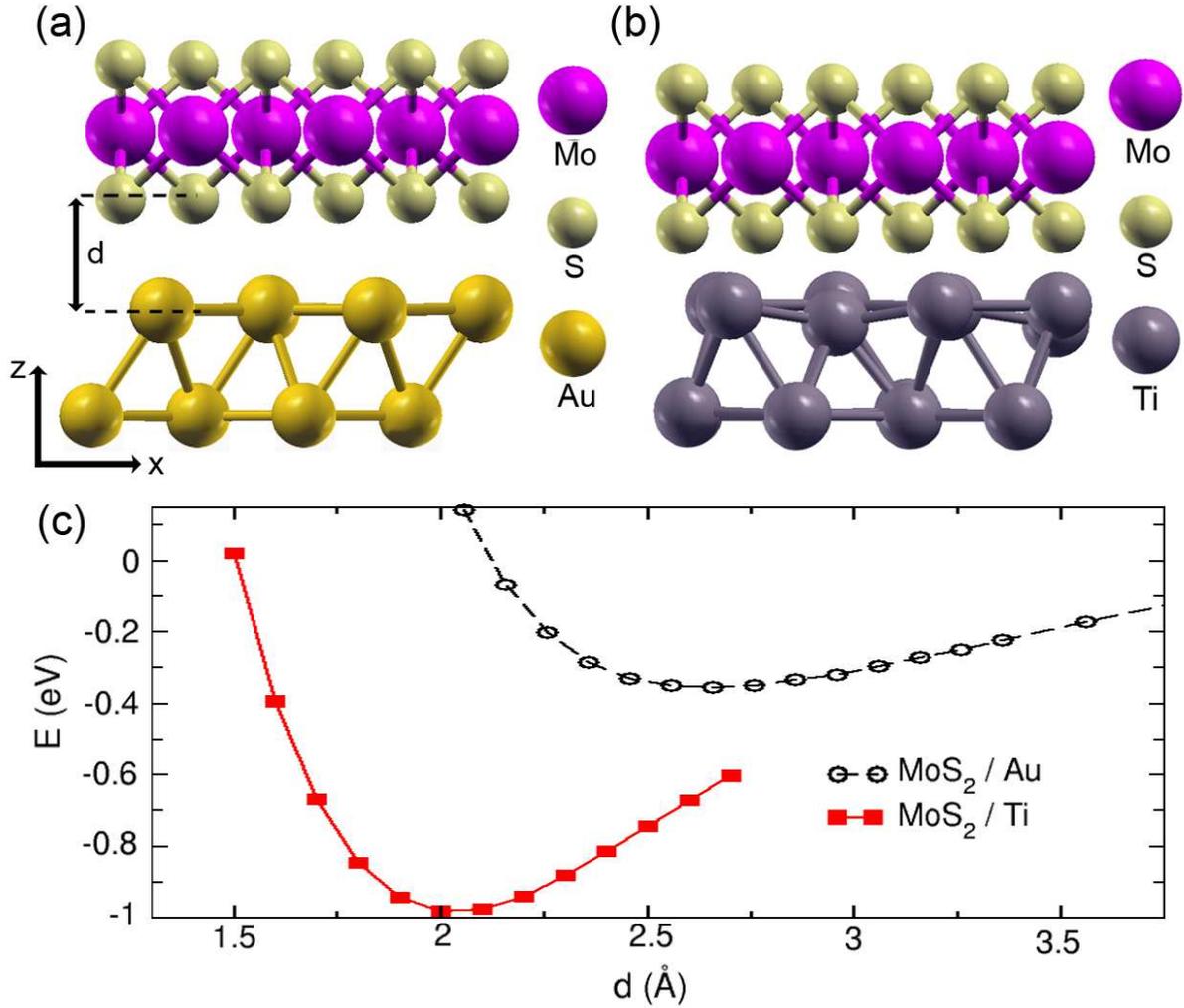}
\caption{(Color online) Side view of the relaxed contact region at
the interface between MoS$_2$ and the (a) Au(111) and (b) Ti(0001)
surface. (c) Binding energy $E$ per interface metal atom as a
function of the separation $d$ between between MoS$_2$ and the
Ti(0001) and Au(111) surface. \label{fig1}}
\end{figure}


The supercell geometry of the relaxed commensurate contact region
between MoS$_2$ and the close-packed surfaces of Au and Ti is
shown in Figs.~\ref{fig1}(a) and \ref{fig1}(b). We have
represented the metal surfaces by 6-layer slabs. The single-layer
MoS$_2$ consists of a molybdenum monolayer sandwiched between two
sulfur monolayers. In each unit cell we distinguish between two
types of S atoms: two S atoms are located in the on-top site and
six S atoms in the hollow site of the metal substrate. When
optimizing this structure using the conjugate gradient technique,
all atoms except the bottom four layers of the metal slabs were
allowed to relax.
In principle, the precise geometry at the metal-semiconductor
interface may be affected by our choice of the
exchange-correlation functional, which does not describe
dispersive interactions accurately. Since the interface bonds are
either semi-covalent or covalent and contain in our estimate only
${\alt}20$\% van der Waals character, we expect the optimized
geometry to be adequate for our study. We noticed that the
relaxation within the MoS$_2$ structure after contacting the
metals was very small.


The following major factors determine the
electronic
transparency of contacts: favorable interface geometry
and bonding, the electronic density of states, and the potential
barrier at the interface. Strong interconnects are especially
important when contacting flexible semiconductors such as MoS$_2$,
which is known to form both planar and tubular
nanostructures\cite{Tenne_1992}. Favorable geometry precludes a
small lattice mismatch at the interface, and should maximize the
overlap between the states at both sides of the interface. The
density of states (DOS) at the Fermi level ($E_F$) should be large
throughout the interface region, forming delocalized states with
low effective electron mass in order to efficiently transfer
electrons between the metal and the semiconductor. The potential
barrier at the interface should be as narrow and low as possible
to maximize current injection. In the following, we analyze each
of these factors for the interface between MoS$_2$ and Au and Ti
as contact metals. As will become clear in the following, Ti turns
out to be superior to the commonly used Au as a contact metal.


Contrary to chemically not saturated sulfur, which forms favorable
thio bonds to Au, the sulfur in MoS$_2$ is fully saturated and
does not bond strongly to Au. This is reflected in the fact that
the shortest distance of 2.62~{\AA} between S in the on-top site
and the Au atoms directly underneath is about 0.2~{\AA} longer
than the sum of the S and Au covalent radii. The distance between
the S atoms in the hollow sites and their closest Au neighbors is
significantly larger, namely 3.15~{\AA}. The average separation
between the top layer of Au and the Mo layer is 4.21~{\AA}, large
enough to suppress any efficient wavefunction overlap.


When compared to Au, the equilibrium separation between Ti and
MoS$_2$ is much lower. In this case, the majority S atoms, which
are in the hollow site, play the key role for adjusting the
interlayer separation, trying to replicate the environment of Ti
in the stable compound TiS$_2$. With the equilibrium distance
between S atoms at the hollow site and its closest Ti neighbors of
2.54~{\AA}, the optimum separation between MoS$_2$ and Ti is about
2.0~{\AA}, much shorter than the sum of the Ti and S covalent
radii of 2.38~{\AA}. The resulting repulsion between the minority
S atoms, which are in the on-top site, pushes away the Ti atoms
underneath, reaching an equilibrium distance of 2.34~{\AA}, close
to the sum of the respective covalent radii.
The average Mo-Ti distance is 3.57~{\AA}, which is 0.64~{\AA}
shorter than the Mo-Au distance, indicating favorable conditions
for a large wavefunction overlap.


To characterize the contact bond strength, we define the binding
energy $E$ between the metal and the MoS$_2$ layer as the total
energy difference between the combined and the isolated systems
and display our results in Fig.~\ref{fig1}(c). We find that the
binding of MoS$_2$ to Au is considerably weaker than to Ti, with
the binding energy per surface metal atom of 0.36~eV in case of Au
as compared to 0.98~eV in case of Ti.

\begin{figure}[tbp!]
\includegraphics[width=0.9\columnwidth]{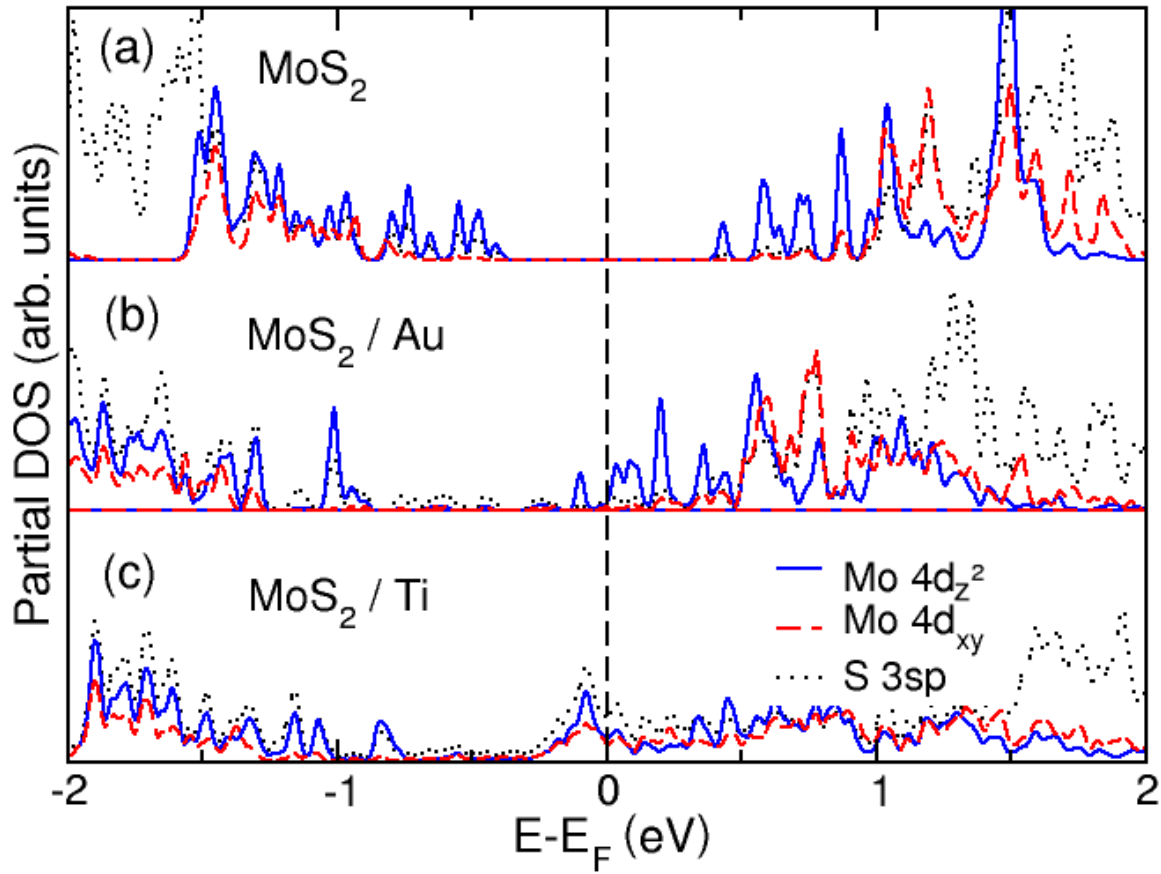}
\caption{(Color online) Partial electronic density of Mo and S
states, which are relevant to bonding and charge injection, in (a)
the single-layer MoS$_2$, (b) the MoS$_2$/Au system, and (c) the
MoS$_2$/Ti system. Only a narrow energy region around the Fermi
level $E_F$ is shown. \label{fig2}}
\end{figure}


The
electronic
transparency of a contact can be quantified in a quantum
transport calculation. For low bias voltages, a suitable approach
may involve calculating the equilibrium Green's function, which to
a large extent reflects the electronic density of states near
$E_F$ and the degree of delocalization of these states within the
contact region. For a detailed insight into the reason, why some
contacts are better than others, we proceed with a careful
analysis of these quantities.


The DOS projection onto selected Mo and S orbitals is presented in
Fig.~\ref{fig2} for the single-layer MoS$_2$, MoS$_2$/Au and
MoS$_2$/Ti. The bottom of the conduction band and the top of the
valence band of the single-layer MoS$_2$ is dominated by
Mo$4d_{z^2}$ states, with the other Mo states playing a minor
role.
Since the work function $\Phi ($MoS$_2)=5.2$~eV is larger than
that of most metals, the
electronic
transparency of the contact is maximized when electronic
states at the Fermi level of the contact metal align and strongly
overlap with the Mo$4d_{z^2}$ states near the bottom of the
conduction band.

\begin{figure}[tbp!]
\includegraphics[width=\columnwidth]{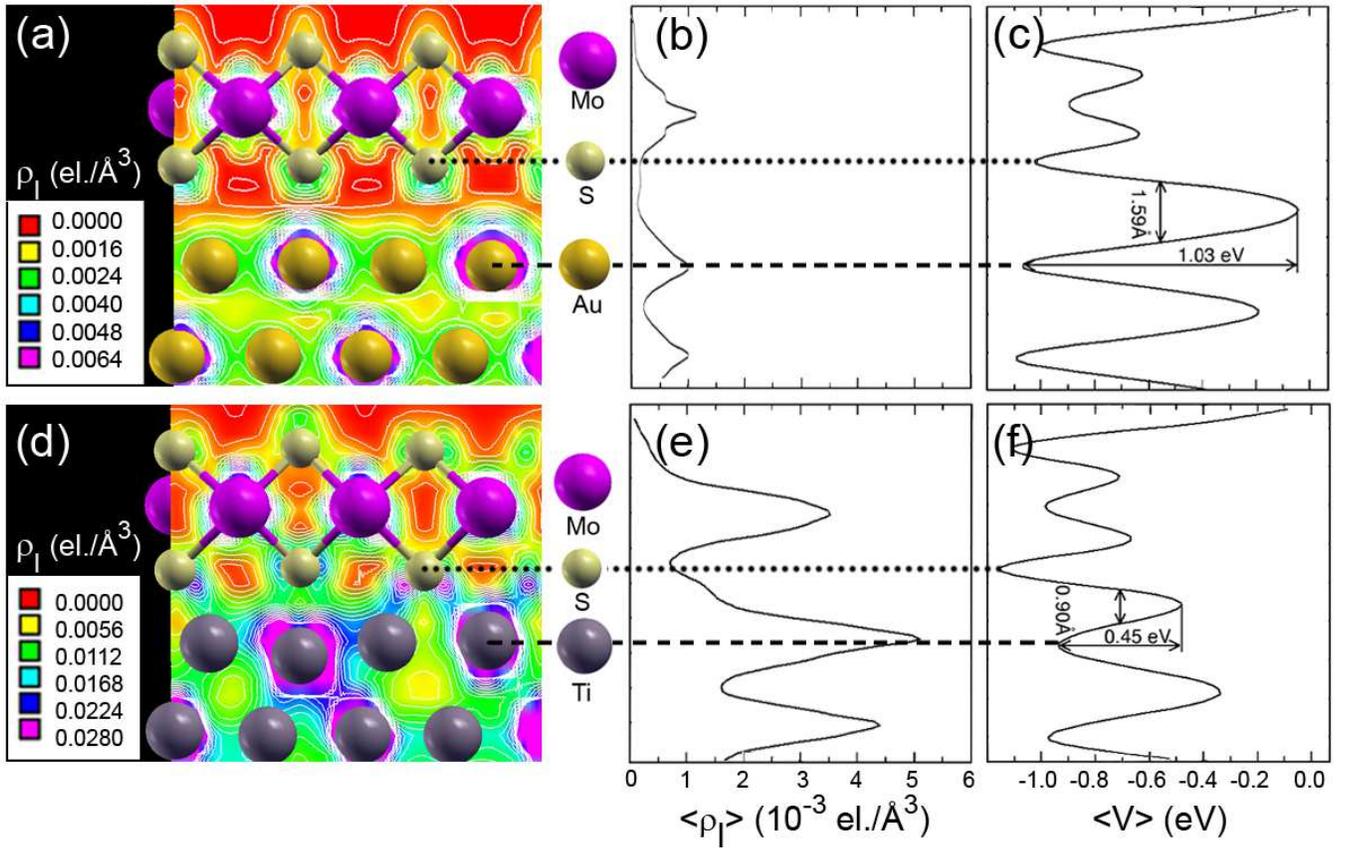}
\caption{(Color online) Electronic structure at the interface
between MoS$_2$ and (a-c) Au and (d-f) Ti. Contour plots of the
charge density $\rho_l$ associated with states in the energy range
$E_F-0.1$~eV$<E<E_F+0.1$~eV in planes normal to the interface in
(a) MoS$_2$/Au and (d) MoS$_2$/Ti. Average value of
$<{{\rho}_l}>(z)$ in planes parallel to the interface of (b)
MoS$_2$/Au and (e) MoS$_2$/Ti. Average electrostatic potential
$<$$V_l$$>(z)$ in planes normal to the interface in (c) MoS$_2$/Au
and (f) MoS$_2$/Ti. The dotted line in the panels indicates the
location of the sulfur layer closest to the metal and the dashed
line the position of the topmost metal layer. \label{fig3} }
\end{figure}


As seen in Fig.~\ref{fig2}(b), upon making contact with Au, the
MoS$_2$/metal interface becomes metallic. The Fermi level of the
combined system shifts upwards, to about 0.1~eV above the bottom
of the conduction band of MoS$_2$. The states near $E_F$ display
dominant Mo$4d_{z^2}$ character with only a small admixture of
S$3sp$ states. Consequently, electron injection into the
semiconductor will involve primarily the Mo$4d_{z^2}$ states that,
according to Fig.~\ref{fig2}(b), display a low partial DOS at
$E_F$.
The corresponding low carrier density near $E_F$ of the interface
is depicted in Fig.~\ref{fig3}(a) and Fig.~\ref{fig3}(b).


As seen by comparing Fig.~\ref{fig2}(b) for MoS$_2$/Au and
Fig.~\ref{fig2}(c) for MoS$_2$/Ti, Ti as contact metal modifies
the electronic states near $E_F$ much more than Au. In the
MoS$_2$/Ti system, the Fermi level is shifted upwards, to 0.25~eV
above the bottom of the MoS$_2$ conduction band. This is much
higher than in the Au system and causes an increase in the DOS at
$E_F$.
The most striking difference to Au is a significant contribution
of S$3sp$ and Mo$4d_{xy}$ states near the Fermi level, which is
associated with a strong S-Ti mixing.
The contribution of Mo$4d_{xy}$ states at $E_F$ is nearly equal to
that of the Mo$4d_{z^2}$ states. The broadening of the peaks in
the DOS near $E_F$ reflects an increase in the dispersion of the
corresponding bands and suggests the formation of delocalized
states at the interface.

The character of the states, which determine the low-bias
transport, is represented by the density $\rho_l({\bf r})$ of the
corresponding carriers in the left panels of Fig.~\ref{fig3}(a)
for MoS$_2$/Au and Fig.~\ref{fig3}(d) for MoS$_2$/Ti.
As seen in Fig.~\ref{fig3}(a), which shows a detailed contour plot
of $\rho_l$, and Fig.~\ref{fig3}(b), representing the average
$<{{\rho}_l}>(z)$ in planes parallel to the interface, the charge
carrier density in the interface region between MoS$_2$ and Au is
very low.
Consequently, the electron transport across the MoS$_2$/Au contact
is mainly of tunneling nature. Since according to
Fig.~\ref{fig2}(b) the electron injection into the MoS$_2$ layer
proceeds exclusively via the Mo$4d_{z^2}$ states, the tunnel
barrier from Au to MoS$_2$ is very wide.

In striking contrast to the MoS$_2$/Au interface, the charge
carrier density in the interface region between MoS$_2$ and Ti is
much larger. This is seen especially when comparing the averaged
carrier density $<{\rho_l}>$ in planes parallel to the interface
in Fig.~\ref{fig3}(b) for MoS$_2$/Au and Fig.~\ref{fig3}(e) for
MoS$_2$/Ti. Of particular interest is the difference between the
electron density at the interfacial sulfur layer, denoted by the
dotted line in Fig.~\ref{fig3}. The nearly vanishing value of
$<{\rho_l}>$ at this location in MoS$_2$/Au increases by an order
of magnitude in MoS$_2$/Ti, thus turning tunneling into resonant
transport. The carrier density delocalization in MoS$_2$/Ti,
anticipated above due to the DOS broadening, corresponds to a
metallization of the interface. This in turn enables direct charge
injection into the MoS$_2$ layer, for which the actual distance
between Mo and Ti atoms becomes irrelevant.


In order to complete the analysis of the contacts, we investigate
the electrostatic potential regarding the existence of
barriers at the metal-semiconductor interface and show the results
in Fig.~\ref{fig3}(c) for MoS$_2$/Au and Fig.~\ref{fig3}(f) for
MoS$_2$/Ti.
Since we observe not only a net charge transfer across the
metal-semiconductor interface, but also changes in the electronic
structure due to the covalent interaction, these barriers are not
ideal Schottky barriers, but rather more general contact tunnel
barriers.
We define the height of the
contact tunnel
barrier as the difference between the averaged potential at the
top metal layer, indicated by the dashed line in Fig.~\ref{fig3},
and the maximum of the averaged potential between the metal and
the neighboring sulfur layer.


As anticipated, the
tunnel
barrier at the interface between the semiconducting MoS$_2$ layer
and the Au surface, which is shown in Fig.~\ref{fig3}(c), is
relatively high (1.03~eV) and wide. Since the S states of the
bottom MoS$_2$ layer are nearly absent near $E_F$ in the
MoS$_2$/Au system, the true barrier is even wider than the
1.59~{\AA} value at half-height shown in Fig.~\ref{fig3}(c). In
this case, tunneling involves direct charge transfer from Au to Mo
states across two barriers.


Electron injection from Ti to MoS$_2$ is completely different. As
seen in Fig.~\ref{fig3}(f), electrons in this system have to
bypass the much lower (0.45~eV) and narrower (0.9~{\AA}) barrier
to reach the delocalized states at the MoS$_2$/Ti interface. In
comparison to Au as contact metal, the significant reduction of
the barriers at the interface with Ti will significantly improve
the
electronic transparency of the contact.


Even though Au is commonly believed to be the ideal contact metal
to many sulfur-terminated systems including multi-layer MoS$_2$,
our study shows that the opposite is true when contacting
single-layer MoS$_2$. A multi-layer system is preferentially
contacted from the side, where Au can bond chemically to not
saturated sulfur atoms at the edge. Since contacting a
single-layer from the side is insufficient for good electron
injection, the preferred geometry is a top contact discussed here.
In this scenario, we identified unexpected qualitative differences
between different contact metals in the way they inject carriers
into MoS$_2$.

The basic difference is that between an inefficient tunnel contact
in MoS$_2$/Au and a low-resistance ohmic contact providing a
direct injection channel in MoS$_2$/Ti. We discuss MoS$_2$/Ti only
as a representative example of an optimum designer contact that is
superior to the state-of-the-art. Good contact metal candidates
must, of course, first fulfill macroscopic criteria such as high
conductivity and chemical, thermal and electrical stability.
Additional criteria for an optimum contact in nanoelectronics,
which we find fulfilled in the case of Ti, include a favorable
interface geometry and bonding. In terms of electronic structure,
an optimum contact has a high density of delocalized states across
the interface at the Fermi level of the combined system,
corresponding to a minimized or non-existent
tunnel
barrier between the two materials.

In conclusion, we performed \textit{ab initio} density functional
theory calculations of MoS$_2$/Au and MoS$_2$/Ti contacts to study
the reason, why single-layer molybdenum disulfide
appears to fall
short of its promising potential in flexible electronics according
to recent experiments\cite{Kis,Novoselov2005}. We found that the
nature of contacts plays a more important role in these systems
than the semiconductor itself. Our calculations for the geometry,
bonding and electronic structure of the contact region suggest
that the most common contact metal (Au) forms a tunnel contact to
single-layer MoS$_2$ and thus is rather inefficient for electron
injection. We find that Ti is a suitable alternative as electrode
material, since it forms a low-resistance ohmic contact. We also
provide specific criteria for selecting materials besides Ti that
should optimize
the electronic transparency of the contact.
Higher contact transparency reduces the required bias voltages for
operation and may also improve the frequency response of these
structurally flexible electronic devices, which may
eventually open new horizons for electronics based on transition
metal chalcogenides.

DT was partly supported by the National Science Foundation
Cooperative Agreement \#EEC-0832785, titled ``NSEC: Center for
High-rate Nanomanufacturing''. GS was partly supported by the
European Research Council (ERC - project INTIF 226639). IP was
supported by the Science Foundation of Ireland (SFI) and CRANN.
Computational resources for this project were provided by the ZIH
Dresden and the Trinity College High Performance Computing Center
(TCHPC).


\begin{thebibliography}{18}
\expandafter\ifx\csname natexlab\endcsname\relax\def\natexlab#1{#1}\fi
\expandafter\ifx\csname bibnamefont\endcsname\relax
  \def\bibnamefont#1{#1}\fi
\expandafter\ifx\csname bibfnamefont\endcsname\relax
  \def\bibfnamefont#1{#1}\fi
\expandafter\ifx\csname citenamefont\endcsname\relax
  \def\citenamefont#1{#1}\fi
\expandafter\ifx\csname url\endcsname\relax
  \def\url#1{\texttt{#1}}\fi
\expandafter\ifx\csname urlprefix\endcsname\relax\def\urlprefix{URL }\fi
\providecommand{\bibinfo}[2]{#2}
\providecommand{\eprint}[2][]{\url{#2}}

\bibitem[{\citenamefont{L\'eonard and Talin}(2006)}]{Leonard2006}
\bibinfo{author}{\bibfnamefont{F.}~\bibnamefont{L\'eonard}} \bibnamefont{and}
  \bibinfo{author}{\bibfnamefont{A.~A.} \bibnamefont{Talin}},
  \bibinfo{journal}{Phys. Rev. Lett.} \textbf{\bibinfo{volume}{97}},
  \bibinfo{pages}{026804} (\bibinfo{year}{2006}).

\bibitem[{\citenamefont{Lin and Jian}(2008)}]{Nanocontact2008}
\bibinfo{author}{\bibfnamefont{Y.-F.} \bibnamefont{Lin}} \bibnamefont{and}
  \bibinfo{author}{\bibfnamefont{W.-B.} \bibnamefont{Jian}},
  \bibinfo{journal}{Nano Lett.} \textbf{\bibinfo{volume}{8}},
  \bibinfo{pages}{3146} (\bibinfo{year}{2008}).

\bibitem[{\citenamefont{Landman et~al.}(2000)\citenamefont{Landman, Barnett,
  Scherbakov, and Avouris}}]{LandmanPRL2000}
\bibinfo{author}{\bibfnamefont{U.}~\bibnamefont{Landman}},
  \bibinfo{author}{\bibfnamefont{R.~N.} \bibnamefont{Barnett}},
  \bibinfo{author}{\bibfnamefont{A.~G.} \bibnamefont{Scherbakov}},
  \bibnamefont{and} \bibinfo{author}{\bibfnamefont{P.}~\bibnamefont{Avouris}},
  \bibinfo{journal}{Phys. Rev. Lett.} \textbf{\bibinfo{volume}{85}},
  \bibinfo{pages}{1958} (\bibinfo{year}{2000}).

\bibitem[{\citenamefont{N.~Nemec and Cuniberti}(2006)}]{DT178}
\bibinfo{author}{\bibfnamefont{D.~T.} \bibnamefont{N.~Nemec}} \bibnamefont{and}
  \bibinfo{author}{\bibfnamefont{G.}~\bibnamefont{Cuniberti}},
  \bibinfo{journal}{Phys. Rev. Lett.} \textbf{\bibinfo{volume}{96}},
  \bibinfo{pages}{076802} (\bibinfo{year}{2006}).

\bibitem[{\citenamefont{Radisavljevic et~al.}(2011)\citenamefont{Radisavljevic,
  Radenovic, Brivio, Giacometti, and Kis}}]{Kis}
\bibinfo{author}{\bibfnamefont{B.}~\bibnamefont{Radisavljevic}},
  \bibinfo{author}{\bibfnamefont{A.}~\bibnamefont{Radenovic}},
  \bibinfo{author}{\bibfnamefont{J.}~\bibnamefont{Brivio}},
  \bibinfo{author}{\bibfnamefont{V.}~\bibnamefont{Giacometti}},
  \bibnamefont{and} \bibinfo{author}{\bibfnamefont{A.}~\bibnamefont{Kis}},
  \bibinfo{journal}{Nature Nanotech.} \textbf{\bibinfo{volume}{6}},
  \bibinfo{pages}{147} (\bibinfo{year}{2011}).

\bibitem[{\citenamefont{Yoon et~al.}(2011)\citenamefont{Yoon, Ganapathi, and
  Salahuddin}}]{Yoon-MoS2-2011}
\bibinfo{author}{\bibfnamefont{Y.}~\bibnamefont{Yoon}},
  \bibinfo{author}{\bibfnamefont{K.}~\bibnamefont{Ganapathi}},
  \bibnamefont{and}
  \bibinfo{author}{\bibfnamefont{S.}~\bibnamefont{Salahuddin}},
  \bibinfo{journal}{Nano Lett.} \textbf{\bibinfo{volume}{11}},
  \bibinfo{pages}{3768} (\bibinfo{year}{2011}).

\bibitem[{\citenamefont{Kam and Parkinson}(1982)}]{MoS2-bandgap}
\bibinfo{author}{\bibfnamefont{K.~K.} \bibnamefont{Kam}} \bibnamefont{and}
  \bibinfo{author}{\bibfnamefont{B.~A.} \bibnamefont{Parkinson}},
  \bibinfo{journal}{J. Phys. Chem.} \textbf{\bibinfo{volume}{86}},
  \bibinfo{pages}{463} (\bibinfo{year}{1982}).

\bibitem[{\citenamefont{Fivaz and Mooser}(1967)}]{Mooser1967}
\bibinfo{author}{\bibfnamefont{R.}~\bibnamefont{Fivaz}} \bibnamefont{and}
  \bibinfo{author}{\bibfnamefont{E.}~\bibnamefont{Mooser}},
  \bibinfo{journal}{Phys. Rev.} \textbf{\bibinfo{volume}{163}},
  \bibinfo{pages}{743} (\bibinfo{year}{1967}).

\bibitem[{\citenamefont{Podzorov et~al.}(2004)\citenamefont{Podzorov,
  Gershenson, Kloc, Zeis, and Bucher}}]{Podzorov}
\bibinfo{author}{\bibfnamefont{V.}~\bibnamefont{Podzorov}},
  \bibinfo{author}{\bibfnamefont{M.~E.} \bibnamefont{Gershenson}},
  \bibinfo{author}{\bibfnamefont{C.}~\bibnamefont{Kloc}},
  \bibinfo{author}{\bibfnamefont{R.}~\bibnamefont{Zeis}}, \bibnamefont{and}
  \bibinfo{author}{\bibfnamefont{E.}~\bibnamefont{Bucher}},
  \bibinfo{journal}{Appl. Phys. Lett.} \textbf{\bibinfo{volume}{84}},
  \bibinfo{pages}{3301} (\bibinfo{year}{2004}).

\bibitem[{\citenamefont{Novoselov et~al.}(2005)\citenamefont{Novoselov, Jiang,
  Schedin, Booth, Khotkevich, Morozov, and Geim}}]{Novoselov2005}
\bibinfo{author}{\bibfnamefont{K.~S.} \bibnamefont{Novoselov}},
  \bibinfo{author}{\bibfnamefont{D.}~\bibnamefont{Jiang}},
  \bibinfo{author}{\bibfnamefont{F.}~\bibnamefont{Schedin}},
  \bibinfo{author}{\bibfnamefont{T.~J.} \bibnamefont{Booth}},
  \bibinfo{author}{\bibfnamefont{V.~V.} \bibnamefont{Khotkevich}},
  \bibinfo{author}{\bibfnamefont{S.~V.} \bibnamefont{Morozov}},
  \bibnamefont{and} \bibinfo{author}{\bibfnamefont{A.~K.} \bibnamefont{Geim}},
  \bibinfo{journal}{Proc. Natl. Acad. Sci. USA} \textbf{\bibinfo{volume}{102}},
  \bibinfo{pages}{10451} (\bibinfo{year}{2005}).

\bibitem[{\citenamefont{Perdew et~al.}(1996)\citenamefont{Perdew, Burke, and
  Ernzerhof}}]{PBE}
\bibinfo{author}{\bibfnamefont{J.~P.} \bibnamefont{Perdew}},
  \bibinfo{author}{\bibfnamefont{K.}~\bibnamefont{Burke}}, \bibnamefont{and}
  \bibinfo{author}{\bibfnamefont{M.}~\bibnamefont{Ernzerhof}},
  \bibinfo{journal}{Phys. Rev. Lett.} \textbf{\bibinfo{volume}{77}},
  \bibinfo{pages}{3865} (\bibinfo{year}{1996}).

\bibitem[{\citenamefont{Soler et~al.}(2002)\citenamefont{Soler, Artacho, Gale,
  Garc\'{i}a, Junquera, Ordej\'{o}n, and S\'{a}nchez-Portal}}]{Soler}
\bibinfo{author}{\bibfnamefont{J.~M.} \bibnamefont{Soler}},
  \bibinfo{author}{\bibfnamefont{E.}~\bibnamefont{Artacho}},
  \bibinfo{author}{\bibfnamefont{J.~D.} \bibnamefont{Gale}},
  \bibinfo{author}{\bibfnamefont{A.}~\bibnamefont{Garc\'{i}a}},
  \bibinfo{author}{\bibfnamefont{J.}~\bibnamefont{Junquera}},
  \bibinfo{author}{\bibfnamefont{P.}~\bibnamefont{Ordej\'{o}n}},
  \bibnamefont{and}
  \bibinfo{author}{\bibfnamefont{D.}~\bibnamefont{S\'{a}nchez-Portal}},
  \bibinfo{journal}{J. Phys. Cond. Matter} \textbf{\bibinfo{volume}{14}},
  \bibinfo{pages}{2745} (\bibinfo{year}{2002}).

\bibitem[{\citenamefont{Popov et~al.}(2007)\citenamefont{Popov, Yang, Berber,
  Seifert, , and Tom\'{a}nek}}]{Popov_PRL}
\bibinfo{author}{\bibfnamefont{I.}~\bibnamefont{Popov}},
  \bibinfo{author}{\bibfnamefont{T.}~\bibnamefont{Yang}},
  \bibinfo{author}{\bibfnamefont{S.}~\bibnamefont{Berber}},
  \bibinfo{author}{\bibfnamefont{G.}~\bibnamefont{Seifert}}, ,
  \bibnamefont{and}
  \bibinfo{author}{\bibfnamefont{D.}~\bibnamefont{Tom\'{a}nek}},
  \bibinfo{journal}{Phys. Rev. Lett.} \textbf{\bibinfo{volume}{99}},
  \bibinfo{pages}{085503} (\bibinfo{year}{2007}).

\bibitem[{\citenamefont{Yang et~al.}(2006)\citenamefont{Yang, Okano, Berber,
  and Tom\'{a}nek}}]{Yang_PRL}
\bibinfo{author}{\bibfnamefont{T.}~\bibnamefont{Yang}},
  \bibinfo{author}{\bibfnamefont{S.}~\bibnamefont{Okano}},
  \bibinfo{author}{\bibfnamefont{S.}~\bibnamefont{Berber}}, \bibnamefont{and}
  \bibinfo{author}{\bibfnamefont{D.}~\bibnamefont{Tom\'{a}nek}},
  \bibinfo{journal}{Phys. Rev. Lett.} \textbf{\bibinfo{volume}{96}},
  \bibinfo{pages}{125502} (\bibinfo{year}{2006}).

\bibitem[{\citenamefont{Popov et~al.}(2008)\citenamefont{Popov, Pecchia, Okano,
  Ranjan, Carlo, and Seifert}}]{Popov_APL}
\bibinfo{author}{\bibfnamefont{I.}~\bibnamefont{Popov}},
  \bibinfo{author}{\bibfnamefont{A.}~\bibnamefont{Pecchia}},
  \bibinfo{author}{\bibfnamefont{S.}~\bibnamefont{Okano}},
  \bibinfo{author}{\bibfnamefont{N.}~\bibnamefont{Ranjan}},
  \bibinfo{author}{\bibfnamefont{A.~D.} \bibnamefont{Carlo}}, \bibnamefont{and}
  \bibinfo{author}{\bibfnamefont{G.}~\bibnamefont{Seifert}},
  \bibinfo{journal}{Appl. Phys. Lett.} \textbf{\bibinfo{volume}{93}},
  \bibinfo{pages}{083115} (\bibinfo{year}{2008}).

\bibitem[{\citenamefont{Troullier and Martins}(1991)}]{Troullier91}
\bibinfo{author}{\bibfnamefont{N.}~\bibnamefont{Troullier}} \bibnamefont{and}
  \bibinfo{author}{\bibfnamefont{J.~L.} \bibnamefont{Martins}},
  \bibinfo{journal}{Phys. Rev. B} \textbf{\bibinfo{volume}{43}},
  \bibinfo{pages}{1993} (\bibinfo{year}{1991}).

\bibitem[{\citenamefont{Artacho et~al.}(1999)\citenamefont{Artacho,
  S\'{a}nchez-Portal, Ordej\'{o}n, Garc\'{\i}a, and Soler}}]{SIESTA_PAO}
\bibinfo{author}{\bibfnamefont{E.}~\bibnamefont{Artacho}},
  \bibinfo{author}{\bibfnamefont{D.}~\bibnamefont{S\'{a}nchez-Portal}},
  \bibinfo{author}{\bibfnamefont{P.}~\bibnamefont{Ordej\'{o}n}},
  \bibinfo{author}{\bibfnamefont{A.}~\bibnamefont{Garc\'{\i}a}},
  \bibnamefont{and} \bibinfo{author}{\bibfnamefont{J.~M.} \bibnamefont{Soler}},
  \bibinfo{journal}{Phys. Stat. Sol. (b)} \textbf{\bibinfo{volume}{215}},
  \bibinfo{pages}{809} (\bibinfo{year}{1999}).

\bibitem[{\citenamefont{Tenne et~al.}(1992)\citenamefont{Tenne, Margulis,
  Genut, and Hodes}}]{Tenne_1992}
\bibinfo{author}{\bibfnamefont{R.}~\bibnamefont{Tenne}},
  \bibinfo{author}{\bibfnamefont{L.}~\bibnamefont{Margulis}},
  \bibinfo{author}{\bibfnamefont{M.}~\bibnamefont{Genut}}, \bibnamefont{and}
  \bibinfo{author}{\bibfnamefont{G.}~\bibnamefont{Hodes}},
  \bibinfo{journal}{Nature (London)} \textbf{\bibinfo{volume}{360}},
  \bibinfo{pages}{444} (\bibinfo{year}{1992}).

\end{thebibliography}


\end{document}